# How a Generation Was Misled About Natural Selection

Subtitle: Natural Selection: How it Works, How it Applies to Culture

Liane Gabora

Department of Psychology, University of British Columbia
Okanagan Campus, Arts Building, 333 University Way, Kelowna BC, V1V 1V7, CANADA

**Abstract**

This article explains how natural selection works and how it has been inappropriately applied to the description of cultural change. It proposes an alternative evolutionary explanation for cultural evolution that describes it in terms of communal exchange.

When science is explained to the general public it is necessary to simplify. Inevitably details get left out, details that some consider important, and the 'sexy' parts of the story get played up. But so long as the overall picture is more or less right, scientists generally appreciate the efforts of popular science writers, the press, and in some cases their fellow colleagues, to make their work accessible to a wider audience. The public in turn benefits from the opportunity to see the world they live in from a new perspective, and consider questions they might not otherwise have considered. Sometimes, though, the baby gets thrown out with the bathwater. The 'babyless' version of a scientific story may be a hit nonetheless. Unless one has advanced training in a highly specialized area of a scientific discipline, it may appear to make sense. In most cases, the misrepresentation of science doesn't make much difference; life goes on as normal. In some cases however, it changes the course of history.

Such is the case with Richard Dawkins' (1976) portrayal of evolution. Dawkins is as successful a popularizer of science as the world has seen. He enchanted a generation with his lucid portrayal of evolution as competition amongst 'selfish genes' and survival of the fittest. His books sparked applications of the theory of natural selection across the social sciences, yielding new sub-disciplines ranging from evolutionary economics, to Darwinian theories of culture and creativity, to neural Darwinism. The allure is understandable. Darwinism is a conceptual framework with considerable explanatory power. Were it possible to root these disciplines in a Darwinian framework, we might arguably have achieved the first step toward a unification of the social sciences comparable with Darwin's unification of the life sciences. But applications of Darwinian thinking outside of biology remain on the fringe of academic inquiry three and a half decades after publication of *The Selfish Gene*. The story of how applications of Darwinism in the social sciences flourished and wilted, and Dawkins' role in this, is one of the most fascinating stories in the history of science.

**How Natural Selection Works**

Let's take a brief look at how evolution works, and then at how Dawkins portrayed it. What necessitated Darwin's theory of how species evolve through natural selection is that traits that are *acquired* by an organism over its lifetime are not transmitted to offspring, and thus not passed down through time. Thus, whereas a rock that smashes stays smashed, a rat whose tail is cut off does not give birth to rats with cut-off tails; the rat lineage loses this trait. So the paradox faced by Darwin and his contemporaries was the following: how do species *accumulate* change when traits acquired over each organism's lifetime are *obliterated?*

Darwin's genius was to explain how species adapt over time despite that new modifications keep getting discarded, by looking from the level of the individual to the level of the *population* of interbreeding individuals. He proposed that part of an organism *is* retained (the part we now call DNA). It is responsible for traits shared by relatives, and it undergoes random change from one generation to the next. Darwin realized that individuals who are better equipped to survive in their given environment leave more offspring (are 'selected'). Thus, although their acquired traits are discarded, their inherited traits (often the traits they were born with) are not. Since random changes to inherited traits that are beneficial cause their bearers to have more offspring, over generations, such traits proliferate at the expense of detrimental ones. Over generations this can lead to substantial change in the distribution of traits across the population as a whole.

We now know that the reason that acquired traits are discarded is that organisms replicate not by mere self-copying, but self-copying using a set of genetic instructions made of DNA. In his 1966 book *Theory of Self-Reproducing Automata,* John von Neumann, one of the most brilliant mathematicians who ever lived, studied the work of early geneticists and worked out the minimal algorithmic structure capable of evolving through natural selection. The result is the gold in the innermost chamber of one of the most fascinating intellectual puzzles in the world. Those who seek to understand natural selection or to apply Darwinian thinking to disciplines outside of biology without ever grasping von Neumann's basic insights are susceptible to all manner of pseudoscientific misadventures.

What von Neumann (1966) showed is that DNA forms a self-assembly code that gets used in two distinct ways. One way it is used involves actively deciphering the instructions to construct a highly self-similar 'variant'. In this case, the code functions as *interpreted information*. The second way is as a self-description that is passively copied to the replicant to ensure that it can reproduce. In this case, the self-assembly instructions function as *un-interpreted information*. (To put it more loosely, the interpreting of the code can be thought of as 'now we make a body', and the un-interpreted use of the code as 'now we make something that can itself make a body'.) Von Neumann refers to a structure that evolves through natural selection as a *self-replicating automaton*.

Von Neumann's book had a profound influence on John Holland, and led to Holland's invention of the *genetic algorithm,* or GA: a computer algorithm that embodies the abstract structure of natural selection (Holland, 1975). The GA solves complex problems by randomly generating variant solutions and selectively 'mating' the best for multiple 'generations' until a satisfactory solution is found. In his 1975 book where the GA was introduced, titled *Adaptation in Natural and Artificial Systems*, Holland proposed three fundamental principles of natural selection. Since biological evolution is the evolutionary process we know the most about, and the process that inspired this analysis, it is easiest to explain these principles in terms of how they apply to biology. The idea, however, is that they apply to *any* system -- biological or otherwise -- that undergoes Darwinian evolution.

The first is *sequestration of inherited information*. The self-assembly code is copied -- without interpretation -- to produce sex cells or *gametes* during meiosis. If gametes unite to form a germ cell, their DNA is decoded-interpreted-to synthesize the proteins necessary to assemble a body. The process by which this unfolds is referred to as development. The DNA self-assembly code, i.e. the information that gets inherited, is said to be *sequestered* because it is encoded in self-assembly instructions and shielded from environmental influence.

The distinction between these two ways of using a self-assembly code, one enabling gamete production and the other enabling development, leads to the second principle: a distinction between genotype and phenotype. The *genotype* of an organism is its genetic constitution: the set of genes it inherited from its parents. The *phenotype* of an organism is its observable characteristics or traits, including morphological, developmental, biochemical, physiological, and behavioral traits. Phenotypes result through the expression of an organism's inherited genotype in the context of its environment. Phenotypes are altered -- they acquire traits -- through interaction with an environment, while genotypes generally do not.

This leads to Holland's third principle: natural selection incorporates not just a means by which inherited variation is *passed on* (*e.g.*, you may have inherited your mother's blue eyes), but also a means by which variation acquired over a lifetime is *discarded* (e.g., you did not inherit your mother's tattoo). What gets transmitted from parent to offspring is sets of self-assembly instructions, and these instruction sets have not changed since they were formed. So traits acquired by parents during their lives, including learned behavior and knowledge, are not inherited by offspring. Since acquired change can accumulate orders of magnitude faster than inherited change, if it is not getting regularly discarded, it quickly overwhelms the population-level mechanism of change identified by Darwin; it 'swamps the phylogenetic signal'. Thus natural selection only works as an explanation when acquired change is negligible.

In sum, building on von Neumann's pioneering insights, Holland identified three basic principle of natural selection: (1) sequestration of inherited information, (2) a clear-cut distinction between genotype and phenotype, and (3) no transmission of acquired traits. Chris Langton, who initiated the field of artificial life, points to the same three principles (Langton, 1992).

**The 'Selfish Gene' View of Evolution**

At the same time as Holland was writing his book, Dawkins was writing *The Selfish Gene* (Dawkins, 1976). In it he took the gene-centric view of evolution developed by George C. Williams, W. D. Hamilton, and John Maynard-Smith, and reformulated it in a way as that sparked the imagination of the general public.

The basic idea is that biology is best understood by looking at the level of not the organism, nor the group, but the gene. A gene, according to Dawkins, is a replicator, which Dawkins defines as "any entity in the universe which interacts with its world, including other replicators, in such a way that copies of itself are made" (Dawkins, 1976; p. 17). A replicator is said to have the following properties:

- *Longevity* - it survives long enough to replicate, or make copies of itself.

- *Fecundity* - at least one version of it can replicate.

- *Fidelity* - even after several generations of replication, it is still almost identical to the original.

Replicators cause the entities that bear them to act in ways that make them proliferate. For example, our genes cause us to act altruistically toward others such as close relatives who share those genes.

That's pretty much it. Note that none of the three basic principles of natural selection identified by Holland play any role in the replicator concept. Note that the replicator concept, unlike the concept of a self-replicating automaton, does not explain why acquired traits (such as your mother's tattoo) are not passed on, but inherited traits (such as your mother's blue eyes) are. Note that the replicator concept does not address the paradox that caused Darwin to come up with the theory of natural selection in the first place: the paradox of how species become increasingly adapted to their environment despite that changes acquired by an organism during its lifetime are not transmitted to offspring.

Dawkins suggested that replicators appear not just in biology but also in culture, and he christened these cultural replicators memes. Memes, like genes, are said to evolve through a Darwinian process. Dawkins writes: "Just as genes propagate themselves in the gene pool by leaping from body to body via sperm or eggs, so memes propagate themselves in the meme pool by leaping from brain to brain."

**Reaction of Biologists**

Other than those who profited from Dawkins' popularization of their ideas, most leading evolutionary biologists, particularly Stephen Jay Gould, Niles Eldredge, Richard Lewontin, Ernst Mayr, Carl Woese, Freeman Dyson, and Stuart Kauffman, were unreceptive to Dawkins' ideas (Dyson, 1999; Fracchia & Lewontin, 1999; Kauffman, 1999; Mayr, 1996; Sterelny, 2007; Temkin & Eldredge, 2007; Woese, 2004). Ernst Mayr, one of the foremost evolutionary biologists of the 20th century, claimed that the replicator notion is "in complete conflict with the basics of Darwinian thought". I once had the interesting experience of driving Ernst Mayr, who was almost 100 years old at the time, from UCLA to a place an hour and a half away. He was charming, but the mere mention of Dawkins unsettled him so much that I thereafter avoided discussion of anything related to him. Stuart Kauffman describes Dawkins' ideas as "impoverished", and claims that the replicator concept does not capture the essential features of the kind of structure that evolves through natural selection.

Incredibly, although theoretical biologists appear to find Dawkins so *wrong* that his ideas are almost unworthy of scientific scrutiny, many social scientists believe him to be so *right* that his ideas don't need scrutiny! Though leading biologists such as Lewontin and Eldredge published papers on the misapplication of Darwinism to culture, by the time they did so the floodgates had already opened, and cultural Darwinism was flourishing (Aunger, 2000; Blackmore, 1999). Social scientists of all stripes were tantalized by the prospect that Darwinism would turn their discipline into a 'real science'. Dawkins also revived interest in evolutionary epistemology, a subfield of philosophy based on the idea that knowledge evolves through a process of conjecture and refutation.

**Memes and the Darwinian Theory of Culture**

Let us now examine the idea of analyzing culture in terms of memes and natural selection. Culture involves the social transmission of novel behavior, such as gestures, songs, or tool-making skills, amongst individuals who are generally members of the same species. Elements of culture spread both vertically, from one generation to another, and horizontally amongst members of a generation. Thus two key components of culture are a means of generating novel behavior, and a means of spreading it, such as imitation and other forms of social learning.

It has been argued that culture constitutes a second evolutionary process, one which, though it grew out of biological evolution, exhibits an evolutionary dynamic in its own right that cannot be reduced to biology. Some of what is considered cultural behavior *can* be explained by biology.

However most would probably concede that, much as principles of physics do not go far toward an explanation of, say, the vertebrate body plan (though things like gravity play some role), biology does not go far toward an explanation of, say, the form and content of a sonnet (though factors like selective pressure for intelligence play some role). Most would concede that to explain how and why such forms arise, accumulate, and adapt over time, one must look to culture. Not only does culture accumulate over time, but it adapts, diversifies, becomes increasingly complex, and exhibits phenomena observed in biological evolution such as niches, and punctuated equilibrium. Like biological evolution, culture is open-ended; there is no apparent limit to the variety of new forms it can give rise to.

So the notion that culture evolves appears to make sense. But the Darwinian approach to culture, and in particular the application of Dawkins' replicator concept, is problematic. The problem is that a Darwinian explanation only works when acquired change is negligible. Otherwise, as explained above, acquired change, which can operate strategically and instantaneously, quickly drowns out inherited change, which operates randomly and only operates at the transition between one generation and the next. Since the theory of natural selection assumes that variation is randomly generated and acquired change is not transmitted, to the extent that transmission is biased from random, and ideas acquire change between transmission events as we contemplate them, a Darwinian framework is inappropriate. (Some apply the term 'Lamarckian' to culture, but having seen that different people interpret this term differently I have come to avoid it.) The only way to maintain a Darwinian perspective on culture that is consistent with the algorithmic structure of natural selection is to view humans as passive imitators and transmitters of prepackaged units of culture, which evolve as separate lineages. To the extent that these lineages 'contaminate one another' -- that is, to the extent that we actively transform elements of culture in ways that reflect our own internal models of the world, altering or combining them to suit our needs, perspectives, or aesthetic sensibilities -- natural selection cannot explain cultural change (Gabora, 2004, 2008, 2011). In order for something to stick in memory we first relate it to what else we know, make it our own, but a Darwinian framework for culture is incompatible with this.

*Dual inheritance theory* is a version of the Darwinian view that posits that humans have two inheritance systems, biological and cultural (Boyd & Richerson, 1985). Proponents of this theory go even further than Dawkins and the 'memeticists'; they say that natural selection does not even require self-replication. They not only throw the baby out with the bathwater, they scour away any of the baby's dead skin cells clinging in the soap scum. Dual inheritance theorists focus on processes that bias the transmission of cultural information, such as the tendency to preferentially imitate high prestige individuals. But since natural selection assumes that variation is randomly generated, to the extent that transmission is biased from random, it is not due to selection, the mechanism Darwin identified; therefore a Darwinian approach actually gives a distorted model. The use of the term dual inheritance to refer to both what is transmitted genetically and what is transmitted culturally is technically incorrect and misleading. That which is transmitted through culture falls under the category of acquired change, not inherited change. Advocates of dual inheritance theory apply phylogenetic methods, developed for classifying biological organisms into lineages, to cultural artifacts. This works well for highly conserved assemblages, but falsely classifies similarity due to horizontal exchange of ideas as similarity originating from a common ancestor.

The limitations of the applications of Darwinism to the social sciences that were inspired by Dawkins' picture of evolution can be seen most clearly by returning to what necessitated the theory of natural selection in the first place. Natural selection, a theory of population-level change, was put forward to account for the paradox that acquired traits are (with few exceptions) not inherited from parent to offspring. Note how exceptional this is. In most domains of human inquiry, change is retained, and successive changes accumulate. If an asteroid crashes into a planet, the planet cannot revert to the state of having not had the asteroid crash into it. But in the biological domain where, if a rat loses its tail the rat's offspring are not born tail-less, how does one explain how change is retained and accumulates? That was the extraordinary paradox Darwin faced, the paradox for which natural selection provided a solution. An analogous paradox does not exist with respect to culture.

**So How *Does* Culture Evolve?**

Dawkins has gone out of his way to emphasize that (1) by 'selfish' he didn't mean that genes are really selfish, they just act as if they were, (2) humans can transcend their 'genetic selfishness'. Nevertheless, using natural selection, a theory based on competition and survival of the fittest, to explain not just biology but also culture is not a promising route to explaining many aspects of human behavior such as cooperation.

It turns out, however, that there is an evolutionary theory that is better equipped to explain these aspects. I will leave detailed discussion of this for another blog, but mention briefly that this other theory of cultural evolution is inspired by recent efforts to understand the structure of the very earliest entities that could be said to have been alive (Gabora, 2006; Vetsigian, Woese, & Goldenfeld, 2007). These earliest life forms were not self-replicating automata. One could say they were replicators, but they did not have genes (selfish or otherwise). They were self-organized metabolic networks that evolved through a sloppy non-Darwinian process involving transmission of acquired traits, referred to National Medal of Science winner Carl Woese and his colleagues as communal exchange.

It has been proposed that, like these earliest life forms, culture evolves through a non-Darwinian process of communal exchange. What evolves through culture is worldviews, the integrated webs of ideas, beliefs, and so forth, that constitute our internal models of the world, and they evolve, as did early life, not through competition and survival of the fittest but through communal transformation of all (Gabora, 1998, 1999, 2000, Gabora & Aerts, 2009). In other words, the assemblage of human worldviews changes over time not because some replicate at the expense of others, as in natural selection, but because of ongoing mutual interaction and modification. Elements of culture such as rituals, customs, and artifacts reflect the states of the worldviews that generate them. The theory is consistent with network-based approaches to modeling trade, artifact lineages, and the social exchange of knowledge and beliefs, as well as with the unexpectedly high degree of cooperativity in human culture.

A cultural Darwinist recently asked me if it were not possible that two processes are operating interactively: the non-Darwinian process that I have focused on, and another process that, even if they are not strictly Darwinian, operate through a selection-like mechanism. The answer is a highly qualified yes, for given that the non-Darwinian process works strategically at the speed of thought, and selection works on randomly generated variation over generations, the first will render the second negligible. In fact many factors in the universe work together to give the patterns of culture we observe, everything from day length to microbes to weather. Science involves isolating a little piece of reality that you think is important to the explanation of a particular phenomenon. The scientific process has a deductive part, which takes smarts but it's pretty straightforward, and an inductive part, which some say takes intuition and creativity. Deciding which aspects of reality to include and leave out of your model is a matter of induction. How one goes about this reveals how thoroughly one has thought through the myriad factors related to the phenomenon being modeled. One could indeed decide to include selection-like processes in a model of culture, and this might produce a more nuanced model. But since selection-like processes operate without strategy, orders of magnitude slower than strategic thinking and communal exchange, this is not a reasonable place to start building a model of culture. I expect one would get more bang for their buck by incorporating weather patterns.

**Concluding Thoughts**

Given that Dawkins introduced the replicator notion after von Neumann introduced the self-replicating automaton, and given that the replicator is a rather misleading cartoon of the original, how did it become so much better known? We will never know, but perhaps it's partly a matter of presentation. The replicator concept is simple, and if you don't think too deeply about it, it appears to have considerable explanatory power. Even the word 'replicator' itself is catchy; the phrase 'self-replicating automaton' seems clunky by comparison.

Personality may have something to do with it too. I once met Dawkins at a conference at Cambridge and spent several hours sitting across the table from him discussing science. In the company of scientists he came across as arrogant, defensive, and unsure of himself. In the media, however, he comes across as quite the opposite: astute, charismatic, and completely believable in his role as an 'Oxford scientist'. I imagine that in such situations he is surrounded by people whose concern is not his science so much as his public image as a scientist. He plays the role to perfection, and has successfully increased public awareness of, and perhaps even passion for, science. By comparison, von Neumann's book on self-replicating automata was written for a mathematical audience, and published posthumously, and von Neumann was more concerned with political ramifications of science than with public awareness of science. Perhaps also, his role in the development of the atomic bomb deters some from probing more deeply into his life and ideas.

Dawkins' latest book, *The God Delusion*, has made even more of a stir than his books on evolution. I find it unfortunate that attention is diverted from the purely scientific arguments to the rather fruitless (in my opinion) debate about science versus creationism. I haven't read the new book, but it came up now and then at a biweekly meeting of some of the world's top scientists and theologians on creativity in the mind, the universe, and nature at Harvard Divinity School, that I was part of in 2009. The impression I got from both scientists and theologians is that he has again crafted a good read, but it is told through a lens that filters out key pieces of the story that would have given it quite another shape. I suppose that once his sights became set on reaching a popular audience, coming up with simple, satisfying stories about how science works became his job. He does what he can, and leaves it to future generations of science writers to scour untapped realms of meme-space and rescue forgotten babies floating in their bathwater.

## Acknowledgments

The author is grateful for funding from the Natural Sciences and Engineering Research Council of Canada.